\documentclass[lettersize,journal]{IEEEtran}

\IEEEoverridecommandlockouts
\normalsize

\usepackage{cite}
\usepackage{url}
 \usepackage{hyperref}
 \hypersetup{colorlinks=true, linkcolor=red, citecolor=purple, urlcolor=blue} 
\usepackage{amssymb}

\usepackage{amsmath}
\usepackage{array}
\usepackage{amsthm}
\allowdisplaybreaks 
\usepackage{autobreak}
\usepackage{mathptmx}
\usepackage{mathtools, cuted}
\usepackage{algpseudocode}
\usepackage[ruled, vlined, linesnumbered]{algorithm2e}
\usepackage{graphicx}

\usepackage{subcaption} 
\usepackage{bbm}
\usepackage{booktabs}
\usepackage{multirow}
\usepackage{makecell}
\usepackage[table,xcdraw,dvipsnames]{xcolor}
\usepackage{colortbl}
\usepackage[most]{tcolorbox}
\usepackage{placeins}
\usepackage{float}
\usepackage{comment}
\usepackage{enumitem}
\usepackage{setspace}

\allowdisplaybreaks

\newcommand{\rs}{\!\!}
\newcolumntype{C}[1]{>{\centering \arraybackslash}p{#1}}

\newcommand{\mquad}{\qquad\qquad\qquad}

\usepackage{acronym}
\acrodef{ml}[ML]{machine learning}
\acrodef{fft}[FFT]{fast fourier transform}
\acrodef{dd}[DD]{double-directional}
\acrodef{fl}[FL]{federated learning}
\acrodef{nmse}[NMSE]{normalized mean squared error}
\acrodef{hfl}[HFL]{hierarchical federated learning}
\acrodef{rl}[RL]{reinforcement learning}
\acrodef{drl}[DRL]{deep reinforcement learning}
\acrodef{cs}[CS]{central server}
\acrodef{sda}[SDA]{spectral domain analysis}
\acrodef{as}[AS]{angular spread}
\acrodef{ds}[DS]{delay spread}
\acrodef{riss}[RISs]{reconfigurable intelligent surfaces}
\acrodef{scm}[SCM]{stochastic channel model}
\acrodef{bilstm}[BiLSTM]{Bidirectional-long short term memory}
\acrodef{dcm}[DCM]{deterministic channel model}
\acrodef{sdcm}[SDCM]{semi-deterministic channel model}
\acrodef{CS3T-unet}[CS3T-UNet]{cross-shaped separated spatial-temporal UNet}
\acrodef{non-WSSUS}[non-WSSUS]{non-Wide sense stationary uncorrelated scattering}
\acrodef{dod}[DoD]{direction of departure}
\acrodef{doa}[DoA]{direction of arrival}
\acrodef{lstm}[LSTM]{long-short term memory}
\acrodef{bilstm}[BiLSTM]{bidirectional long-short term memory}
\acrodef{gru}[GRU]{gated recurrent unit}
\acrodef{cnn}[CNN]{convolutional neural network}
\acrodef{gnn}[GNN]{graph neural network}
\acrodef{ga}[GA]{ground air}
\acrodef{bs}[BS]{base station}
\acrodef{isp}[ISP]{(wireless) internet service provider}
\acrodef{ue}[UE]{user equipment}
\acrodef{es}[ES]{edge server}
\acrodef{csp}[CSP]{content service provider}
\acrodef{fedavg}[{\tt FedAvg}]{federated averaging}
\acrodef{fednova}[{\tt FedNova}]{federated normalized averaging}
\acrodef{afa}[{\tt AFA}]{anarchic federated averaging}
\acrodef{afacd}[{\tt AFA-CD}]{AFA-cross-device}
\acrodef{feddisco}[{\tt FedDisco}]{federated learning with discrepancy-aware collaboration}
\acrodef{scaffold}[{\tt SCAFFOLD}]{stochastic controlled averaging algorithm}
\acrodef{osafl}[{\tt OSAFL}]{\underline{\textbf{o}}nline-\underline{\textbf{s}}core-\underline{\textbf{a}}ided \underline{\textbf{f}}ederated \underline{\textbf{l}}earning}
\acrodef{iot}[IoT]{Internet of Things}
\acrodef{os}[OS]{operating system}
\acrodef{iid}[IID]{independent and identically distributed}
\acrodef{sca}[SCA]{successive convex approximation}
\acrodef{sgd}[{\tt SGD}]{stochastic gradient descent}
\acrodef{cpu}[CPU]{central processing unit}
\acrodef{gpu}[GPU]{graphics processing unit}
\acrodef{prb}[pRB]{physical resource block}
\acrodef{snr}[SNR]{signal-to-noise-ratio}
\acrodef{lp}[LP]{linear programming}
\acrodef{fpp}[FPP]{floating point precision}
\acrodef{cdf}[CDF]{cumulative distribution function}
\acrodef{uav}[UAV]{unmanned aerial vehicles}
\acrodef{ap}[AP]{access point}
\acrodef{hz}[Hz]{hertz}
\acrodef{csi}[CSI]{channel state information}
\acrodef{gbscm}[GBSCM]{geometry-based stochastic channel model}
\acrodef{kkt}[KKT]{Karush–Kuhn–Tucker}
\acrodef{fifo}[{\tt FIFO}]{first-in-first-out}
\acrodef{trimtoplabel}[{\tt TrimTopLabel}]{trim top label}
\acrodef{fcn}[{\tt FCN}]{fully connected neural network}
\acrodef{lstm}[{\tt LSTM}]{long short-term memory}
\acrodef{bilstm}[{\tt BiLSTM}]{bidirectional long short-term memory}
\acrodef{fc}[{\tt FC}]{fully connected}
\acrodef{isac}[ISAC]{integrated sensing and communication}
\acrodef{mfedavg}[{\tt M-FedAvg}]{modified-FedAvg}
\acrodef{mfedprox}[{\tt M-FedProx}]{modified-FedProx}
\acrodef{mfednova}[{\tt M-FedNova}]{modified-FedNova}
\acrodef{mafacd}[{\tt M-AFA-CD}]{modified-AFA-CD}
\acrodef{mfeddisco}[{\tt M-FedDisco}]{modified-FedDisco}
\acrodef{mpc}[MPC]{multi-path component}
\acrodef{io}[IO]{interacting object}
\acrodef{tx}[TX]{transmitter}
\acrodef{rx}[RX]{receiver}
\acrodef{as}[AS]{angular spread}
\acrodef{ds}[DS]{delay spread}
\acrodef{gcn}[GCN]{graph convolutional network}
 \acrodef{cgan}[CGAN]{conditional generative adversarial network}
\acrodef{ts}[TS]{time series}
\acrodef{knn}[KNN]{K-nearest neighbors}
\acrodef{cdf}[CDF]{cumulative distribution function}
\acrodef{tt}[TNTF]{TimesNet-TimeFilter}
\acrodef{gai}[GAI]{generative artificial intelligence}
\acrodef{gan}[GAN]{generative adversarial network}
\acrodef{llm}[LLM]{large language model}
\acrodef{mMIMO}[mMIMO]{massive MIMO}
\acrodef{wgan-gp}[WGAN-GP]{wasserstein generative adversarial network with gradient penalty}
\acrodef{moe}[MoE]{mixture-of-expert}

\title{Double-Directional Wireless Channel Modeling Using Statistics-Aided Machine Learning }

\author{
\IEEEauthorblockN{Richmond Boamah and Ferdous Pervej}  \\
\IEEEauthorblockA{
Department of Electrical and Computer Engineering, Utah State University, Logan, UT $84332$ \\
E-mails: {\tt A02488524@aggies.usu.edu, ferdous.pervej@usu.edu}} 
\vspace{-0.25in}
}
\begin{document}
\maketitle
\IEEEpeerreviewmaketitle
\thispagestyle{empty}

\setstretch{0.97}

\begin{abstract}
The \ac{dd} wireless channel model is important for realistic system design since it provides complete propagation information. 
While stochastic and deterministic channel models are widely adopted, and existing \ac{ml} solutions mostly aim to align future channel realizations, these solutions are often limited to short time spans that may not be statistically significant.
Moreover, because the number of \acp{mpc} varies with spatial and temporal variation of the \ac{rx} and/or \acp{io}, typical \ac{ml} solutions that require fixed, predefined input and output shapes fall short.
To curb these limitations, we propose a statistics-aided \ac{ml} solution that relies on a fixed subset of \acp{mpc} selection.
More specifically, we first select top-$M$ \acp{mpc}, where $M\in \mathbb{Z}^+$ is much smaller than the total number of \acp{mpc}, and construct learnable graphs to train our proposed hybrid \ac{tt} model.
We then use a channel statistics-aided training method to generate future top-$M$ \ac{dd} channel realizations such that the statistics calculated from these realizations matches closely with those of the actual statistics from the complete time-varying \ac{dd} channel realizations. 
We validate the proposed solution using extensive simulations on both synthetic \ac{scm}-based and deterministic ray-tracing-based datasets, and demonstrate its effectiveness relative to state-of-the-art baselines.
\end{abstract}


\acresetall

\section{Introduction}
\noindent
A communication system depends on the ``channel" it operates on, which is one of the key reasons to call the channel the ``heart" of a communication system.   
In wireless communication, the {\em \ac{dd} channel model} \cite{steinbauer2001dd} is widely used to accurately represent the propagation path. 
Traditionally, {\em stochastic} and {\em deterministic} channel models are widely used \cite{pervej2025double}: stochastic models consider a large-scale generic case that can be applied to various scenarios that are often used for standardization, while the deterministic models use the exact geometry of the environment and are often used for deployment planning.

While stochastic modeling often ignores the exact geometry of the environment and, hence, typically does not accurately model the channel for a given environment, accurate channel modeling for every environment using deterministic ray tracing is highly computationally infeasible. 
As such, we need \ac{ml}-based solutions that can generalize in various propagation environments.

However, because the \ac{dd} channel model provides rich representations of each \ac{mpc} generated by \acp{io} in the environment, naive use of \ac{ml} models to predict the channel realizations may not be useful in a dynamic environment with many \acp{mpc}.
Among the existing \ac{ml}-based channel modeling approaches, predictive modeling \cite{wu2026high}, which matches channel {\em realizations}, and generative modeling \cite{yang2019generative}, which generates {\em synthetic} channel realizations that closely match the actual channel, are widely used.
Nevertheless, when it comes to matching the {\em channel statistics}, these trivial solutions often have limitations (see \cite{pervej2025double} and the references therein).
It has been shown in \cite{pervej2025double,pervej2026dd} that statistical information extracted from historical channel realizations can be directly used to train the model more accurately to generate complete \ac{dd} channel realizations that closely match the ground-truth statistics.

The literature in channel modeling with typical predictive and generative \ac{ml} approaches \cite{lee2024generative} is quite ripe. 
Predictive approaches are generally good for instantaneous channel realization prediction. 
In this predictive direction, \cite{zhou2024transformer} predicted \ac{csi} using a Transformer \cite{vaswani2017attention}, which has quadratic complexity in computing self-attention, potentially becoming computationally costly for long input sequences.
In \cite{kang2024cross}, a cross-shaped separated spatial-temporal UNet model, which has a modified Transformer architecture, was used for
massive multiple-input multiple-output channel prediction. 
Generally, predictive solutions are good at predicting channel realizations over short periods (e.g., a few milliseconds or radio frames), which may not yield meaningful statistics.

In the generative channel modeling paradigm, \cite{cui2025generative} used a \ac{gan} \cite{goodfellow2014generative} to 
generate synthetic channel realizations across distributed users experiencing different channel behaviors in different propagation environments.
In \cite{yang2019generative}, a \ac{gai} model was used to generate \ac{csi} by using raw measurement data. 
\cite{tian2024generative}   proposed a \ac{cgan} model with Wasserstein generative adversarial network-gradient penalty loss
and 1D convolutional neural network block for generating channel parameters, which are conditioned by distinct positions of Tx and Rx.  The \ac{cgan} adds constraints to the generator to produce data based on certain conditions. 

Recently, \cite{pervej2025double, pervej2026dd} proposed statistics-aided channel modeling using a hybrid Transformer model.
While these early works lay the foundations for this research, \cite{pervej2025double,pervej2026dd} considered only a simple \ac{gbscm} based channel model with a fixed set of \acp{mpc} at all trajectory points.
However, in a practical propagation environment, neither the total \acp{mpc} nor any of them may remain active in all future time steps (or trajectory points): \acp{mpc} exhibit a \emph{birth-death} process.

Since the reliability of statistics-aided methods depends on the sample size, we naturally require the ability to generate or predict a long sequence of channel realizations from historical information.
As such, this problem is well-suited to time series forecasting, and thus models such as RNNs, Transformers, and hybrid architectures can be used as the ML model. 
Nevertheless, there are alternative model architectures (e.g., SCINet \cite{liu2022scinet}, TimesNet \cite{wu2022timesnet}) that extract key features using convolutional operations and have achieved significantly better performance than the Transformer.  
Besides, since wireless networks are essentially distributed {\em graphs} \cite{shen2022graph}, {\em \ac{gnn}} \cite{wu2020comprehensive} offers opportunities to leverage domain knowledge to construct the graph from the training data.  

Motivated by the above facts, we present a new {\em graph-based} hybrid \ac{ml} model that requires graph data constructed from propagation characteristics and is trained using \ac{dd} channel statistics to generate a long sequence of complete \ac{dd} channel realizations with accurate statistics.
Our key contributions are summarized below:
   \begin{itemize}[leftmargin=10pt]
   \item Since complete \ac{dd} channel realizations do not have a fixed number of \acp{mpc}, but any \ac{ml} model requires fixed, predefined input/output shape, we extract (a subset) top-$M$ ($M \in \mathbb{Z}^+$) \acp{mpc}, which are much smaller than the total \acp{mpc}, from \ac{dd} channel realizations, followed by using an innovative propagation-knowledge-based patching to construct trainable graph data using Manhattan distance, and {\em temporal}, {\em spatial}, and {\em temporal-spatial} correlations among the patches.
   \item We then design a hybrid \ac{tt} model that combines the benefits of the state-of-the-art TimeFilter \cite{hu2025timefilter} and TimesNet \cite{wu2022timesnet} models to efficiently learn temporal and spatial correlations of the \ac{mpc} features of the \ac{dd} channel. 
    \item To this end, we used a {\em statistics-aided} training method to match the statistics of the future top-$M$ generated \ac{dd} channel realizations to those of the actual statistics calculated from all future \acp{mpc}. 
    Finally, we validate our proposed solution using (a) \ac{scm}-based synthetic datasets and (b) extensive ray tracing datasets for different $M$, historical lags, and prediction lengths. 
     
    \end{itemize}


\section {System Model}
\noindent
We consider a generic system model with a single \ac{tx}, located at a fixed position 
$\tilde{\mathbf{r}}_{\text{tx}} =\{ q_{x,\tilde{\mathbf{r}}}, q_{y,\tilde{\mathbf{r}}}, q_{z,\tilde{\mathbf{r}}}\}$, where $q_{x,\tilde{\mathbf{r}}}, q_{y,\tilde{\mathbf{r}}}, q_{z,\tilde{\mathbf{r}}}$, are the $(x,y,z)$ coordinates of the \ac{tx} operating with an \emph{omnidirectional} antenna and $\tilde{\mathbf{r}}$ is a shorthand notation for $\tilde{\mathbf{r}}_\mathrm{tx}$. 
Given the location of a \ac{rx}, also equipped with an omnidirectional antenna, at 
$\mathbf{r}_{\text{rx}} = \{q_{x,\mathbf{r}}, q_{y,\mathbf{r}}, q_{z,\mathbf{r}}\}$, we express the \ac{dd} channel impulse response as\footnote{The dependency of $\Omega$, $\Psi$, $\tau$, and $a$ on time $t$ is not explicit in this model.} \cite{steinbauer2001dd}
\begin{align}
\label{eq:dd_cir}
h(t,\tau,\Omega,\Psi;
\tilde{\mathbf{r}},\mathbf{r}^l)
  &= \sum\nolimits_{n=1}^{N({\mathbf{r}^l})} |g_{\mathbf{r}^l},n|e^{j\phi_n}
  \delta(\tau - \tau_{\mathbf{r}^l,n})\,
  \delta(\Omega - \Omega_{\mathbf{r}^l,n}) \nonumber\\
  &\mquad \delta(\Psi - \Psi_{\mathbf{r}^l,n})\,
  e^{j2\pi\nu_{\mathbf{r}^l,n} t}, 
\end{align}
where $t, \tau, \Omega, \Psi$ are the time, delay, \ac{dod}, and \ac{doa}, respectively. 
Besides, $N(\mathbf{r})$ is the number of \acp{mpc} at the given 
$(\tilde{\mathbf{r}}_\mathrm{tx}, \mathbf{r}_\mathrm{rx})$. 
Furthermore, $g_n$, $\phi_n$, $\tau_n$, 
$\Omega_n$, $\Psi_n$, and $\nu_n$ are the gain, (random) phase, delay, \ac{dod}, \ac{doa}, and Doppler shift of the 
$n^\mathrm{th}$ path, respectively.

\noindent
We stress that we considered omnidirectional antennas to obtain (\ref{eq:dd_cir}) instead of taking the transfer function, because one can easily generate the transfer function from (\ref{eq:dd_cir}) by multiplying the antenna pattern \cite[Ch. $8$]{molisch2023wireless}, and we are interested in knowing the statistics constructed from the actual propagation channel. 
Besides, this paper considers the \ac{ds} and \ac{as} constructed from the \ac{dd} channel realizations.  
We note that \ac{ds} demonstrates how the \acp{mpc} are spread out in time, while \ac{as} represents their separation in spatial angles. 
Second-order statistics will be considered in our future work.

\subsection{Problem Statement}

\noindent
From (\ref{eq:dd_cir}), one can construct the complete \ac{dd} channel information at a given $\mathbf{r}^l$ as
\begin{align}
{\mathbf{x}_{\mathbf{r}^l}}
&\coloneqq
\left\{ {q}_{x,\mathbf{r}^l}, {q}_{y,\mathbf{r}^l}, g_{\mathbf{r}^l}, d_{\mathbf{r}^l}, \left\{\mathbf{z}_n \right\}_{n=1}^{N(\mathbf{r})}
\right\},
\end{align}
where $\mathbf{z}_n \coloneqq \{ n_{\mathbf{r}^l}, g_{\mathbf{r}^l,n}, \tau_{\mathbf{r}^l,n}, \Omega_{\mathbf{r}^l,\mathrm{az},n}, \Omega_{\mathbf{r}^l,\mathrm{zn},n}, \Psi_{\mathbf{r}^l,\mathrm{az},n}, \Psi_{\mathbf{r}^l,\mathrm{zn},n}\}$.
Besides, $g_{\mathbf{r}^l} := 10 \log_{10} \left( \sum_{n=1}^{N(\mathbf{r})} 10^{\frac{g_{\mathbf{r},n}}{10}} \right)$ is the total gain at the $\mathbf{r}^\mathrm{th}$ receiver position and $d_{\mathbf{r}^l}$ is the distance from transmitter to receiver.
However, since $N(\mathbf{r}^l) \neq N({\mathbf{r}^l}')$, different \ac{dd} channel realizations contain a different number of elements, which is a problem for any \ac{ml} models that typically need a predetermined input and output shape.

To solve this problem, we consider selecting only a subset of $N(\mathbf{r}^l)$ \acp{mpc}.
Namely, we select the top-$M$, where $M \ll N(\mathbf{r}^l)$, \acp{mpc} and construct an approximate channel information tensor as
\begin{align}
\tilde{\mathbf{x}}_{\mathbf{r}^l} 
&\coloneqq \left\{{q}_{x,{\mathbf{r}^l}}, {q}_{y,\mathbf{r}^l}, g_{\mathbf{r}^l}, d_{\mathbf{r}^l},\left\{ \tilde{\mathbf{z}}_{n'}    \right\}_{n'=1}^{M}
\right\}, 
\end{align}
where $\tilde{\mathbf{z}}_{n'} \coloneqq \rs \{n_{\mathbf{r}^l}, g_{{\mathbf{r}^l},n'}, \tau_{{\mathbf{r}^l},n'}, \Omega_{{\mathbf{r}^l}\mathrm{az},n'}, \Omega_{,{\mathbf{r}^l},\mathrm{zn},n'}, \rs \Psi_{{\mathbf{r}^l},\mathrm{az},n'}, \rs \Psi_{{\mathbf{r}^l},\mathrm{zn},n'} \}$.
From $\tilde{\mathbf{x}}_{\mathbf{r}^l}$ we construct the historical \ac{dd} channel sequence $\tilde{\mathbf{X}}_L \coloneqq \left\{\tilde{\mathbf{x}}_1, \dots, \tilde{\mathbf{x}}_{L}\right\} \in \mathbb{R}^{L \times (4 + 7M)}$, where $L \in \mathbb{Z}^+$ is the lag size.
Our goal is to generate the future $P$ top-$M$ \ac{dd} channel realizations  $\hat{\mathbf{X}}_P \coloneqq \left\{\tilde{\mathbf{x}}_{L+1},  \dots, \tilde{\mathbf{x}}_{L+P}\right\}\in \mathbb{R}^{P \times (4 + 7M)}$, such that the statistics of these $\hat{\mathbf{X}}_P$ samples matches closely with those of the true statistics calculated from $\mathbf{X}_P \coloneqq 
\left\{\mathbf{x}_{L+1}, \dots, \mathbf{x}_{L+P}\right\} \in \mathbb{R}^{P \times (4 + 7M)}.$

Note that we calculate the \ac{ds} $s_\tau$ from a true \ac{dd} channel realization as
\begin{equation}
\label{eq:rms_delay}
    s_{\tau} = \sqrt{ \big[ \sum\nolimits_{n=1}^{N(\mathbf{r}^l)} |g_{\mathbf{r}^l,n}|^2 \left(\tau_{\mathbf{r}^l,n}-\bar{\tau}\right)^2 \big] / \big( \sum\nolimits_{n'=1}^{N(\mathbf{r}^l)} |g_{\mathbf{r}^l,n'}|^2 \big) }, 
\end{equation}
where $\bar{\tau} = \big[\sum_{n=1}^{N(\mathbf{r})}
|g_{\mathbf{r}^l,n}|^2\tau_{\mathbf{r}^l,n} \big] / \big[ \sum_{n'=1}^{N^{(\mathbf{r})}} |g_{\mathbf{r}^l,n'}|^2 \big]$.
Besides, the \ac{ds} from the top-$M$ \acp{mpc} is calculated similarly with changing the summation limit to $M$ in \eqref{eq:rms_delay}.

We calculate the \ac{as} from a true realization using the Fleury definition \cite{fleury2002first}, which avoids the ambiguity arising from the periodicity of the azimuthal angle \cite[Ch. $6$]{molisch2023wireless}, as
\begin{equation}
   \rs s_{\Psi} \rs =\rs \sqrt{\rs \big[ \sum\nolimits_{n=1}^{N(\mathbf{r}^l)} | \exp(j\Psi_{\mathbf{r}^l,n})-\mu_{\Psi} |^2 |g_{\mathbf{r}^l,n}|^2 \big] / \big[ \sum\nolimits_{n'=1}^{N(\mathbf{r}^l)} |g_{\mathbf{r}^l,n'}|^2 \big]}, \rs 
\label{eq:rms_angular} 
\end{equation}
where $\mu_{\Psi} = \big[\sum\nolimits_{n=1}^{N(\mathbf{r}^l)} \exp\left(j\Psi_{\mathbf{r}^l,n}\right)\,|g_{\mathbf{r}^l,n}|^2 \big]/ \big[\sum_{n'=1}^{N(\mathbf{r}^l)} |g_{\mathbf{r}^l,n'}|^2\big]$.



\section{Proposed Statistics-Aided Machine Learning Solution}
\noindent 
To solve the above problem, we propose a hybrid \ac{tt} model that requires graph training data and uses a statistics-aided training method.
As such, this section presents our solution in the following order: (i) data preparation stage, which is needed to construct the graph using parameter-specific patching, (ii) proposed \ac{tt} model architecture, which combines the benefits of TimeFilter and TimesNet, and (iii) statistics-aided training method, which uses the statistics of the channel to make the training process efficient to generate the top-$P$ realizations to match the true \ac{dd} channel statistics.

\begin{figure}[!t]
\centering
\includegraphics[width=0.5\textwidth, height=0.6\textheight]{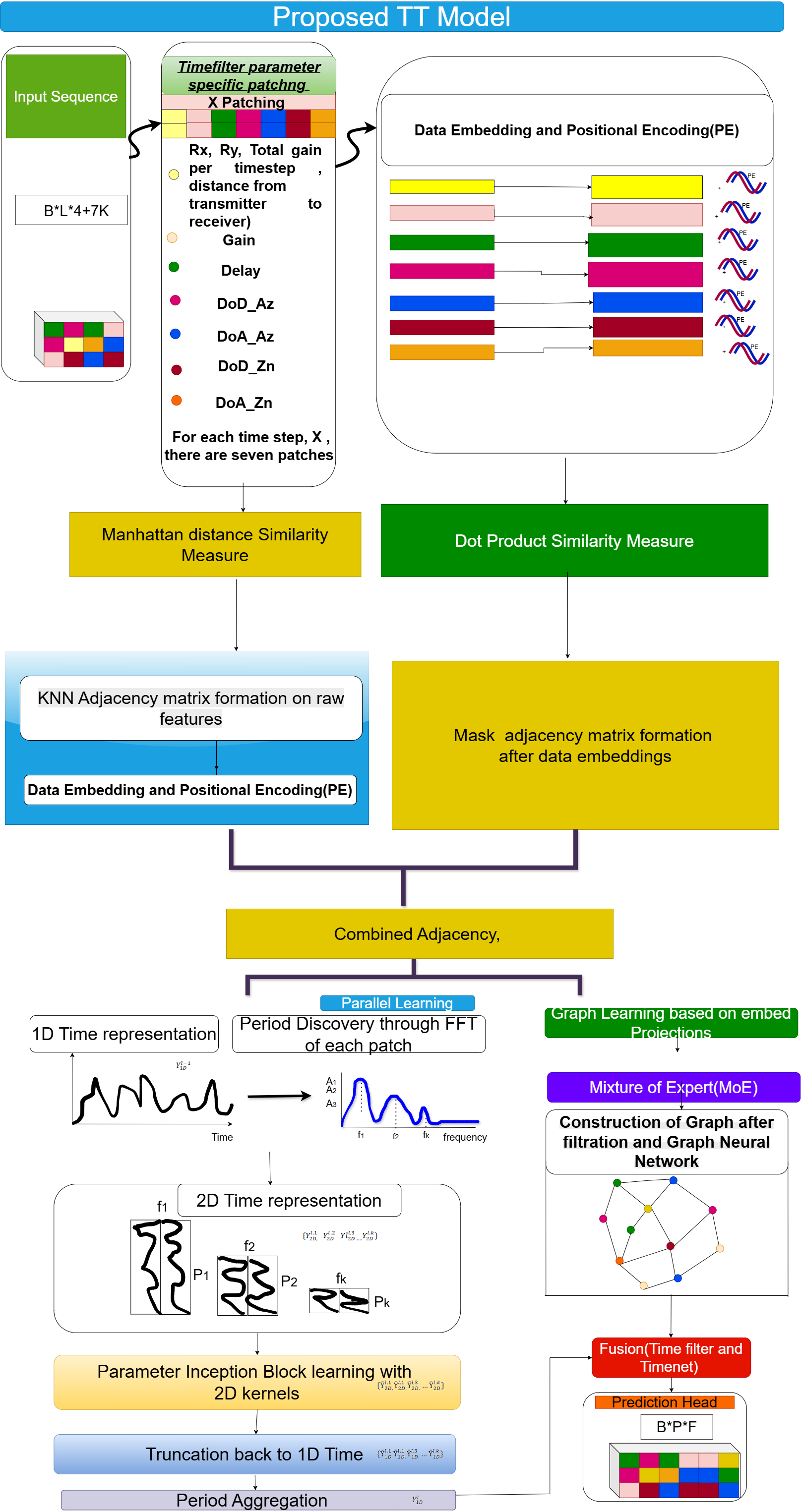}
\caption{Overview of the proposed solution: conversion of \ac{dd} channel realizations into learning graph, followed by transforming node representation into higher-dimensional embedding and transformation for training the proposed \ac{tt} model with a  statistics-aided loss function}
\label{fig:hybridModel}
\end{figure}

\subsection{Data  Preparation}
\noindent
We first leverage wireless propagation knowledge to perform a novel parameter-specific patching, then calculate Manhattan distances among the patches to identify similarities and construct an adjacency matrix, which is then utilized by the graph filter adjacency for graph representation.
These steps are summarized below. 

\subsubsection{Step 1: Parameter-specific Patching} 

\noindent
Since distinct channel parameters have different physical interpretations that can be categorized by the types of \ac{mpc} parameters, we seek a {\em propagation-knowledge-informed} patching strategy for constructing the learnable graph rather than a non-overlapping, sequential patching approach.
Our idea is to first group the \ac{mpc} features by type, then learn the spatial, temporal, and spatial-temporal relationships among them, thereby  
reducing feature interference and making it easier for the model to learn the intricate relationship across features. More specifically, we take the approximate channel sequence $\tilde{\mathbf{X}}_L$ and construct the patches as follows:
$\label{eq:patching}
    \mathcal{X}_{\mathbf{r}^l} \coloneqq \{ \mathbf{p}_{\mathbf{r}^l}^{i}\}_{i=1}^{I=7}$,
where $i=1,2,\dots, 7$, $\mathbf{p}_{\mathbf{r}^l}^{1} \coloneqq\{q_{x,\mathbf{r}^l},q_{y,\mathbf{r}^l}, g_{\mathbf{r}^l}, d_{\mathbf{r}^l}\}$, $\mathbf{p}_{\mathbf{r}^l}^{2} \coloneqq\{g_{\mathbf{r}^l}^{1}, \dots,g_{\mathbf{r}^l}^{M}\}$, $\mathbf{p}_{\mathbf{r}^l}^{3} \coloneqq\{\tau_{\mathbf{r}^l}^{1},\dots,\tau_{\mathbf{r}^l}^{M}\}, $
$\mathbf{p}_{\mathbf{r}^l}^{4}\coloneqq\{\Omega_{\mathrm{az},\mathbf{r}^l}^{1},\dots,\Omega_{\mathrm{az},\mathbf{r}^l}^{M}\}, 
\mathbf{p}_{\mathbf{r}^l}^{5} \coloneqq\{\Omega_{\mathrm{zn},\mathbf{r}^l}^{1},\dots,\Omega_{\mathrm{zn},\mathbf{r}^l}^{M}\}, 
\mathbf{p}_{\mathbf{r}^l}^{6} \coloneqq\{\Psi_{\mathrm{az},\mathbf{r}^l}^{1},\dots,\Psi_{\mathrm{az},\mathbf{r}^l}^{M}\}, \mathbf{p}_{\mathbf{r}^l}^{7} \coloneqq\{\Psi_{\mathrm{zn},\mathbf{r}^l}^{1},\dots,\Psi_{\mathrm{zn},\mathbf{r}^l}^{M}\}$.

\subsubsection{Step 2: Manhattan Distance}

\noindent
Next, we measure similarities among the extracted patches to learn the interrelationships among the various time steps (i.e., trajectory points) in their original forms. 
More specifically, we use the Manhattan distance, i.e., $L_1$ metric, which is computed for each patch's every time steps to construct the distance matrix $\mathbf{D}$.
That is, for the $i^\mathrm{th}$ patch in $\mathcal{X}_{\mathbf{r}^l}$ and $\mathcal{X}_{{\mathbf{r}^l}'}$, where $\mathbf{r}^l \neq \mathbf{r}^l$, this $L_1$ distance is  
\begin{equation}
    {\mathbf{D}}[i, {\mathbf{r}^l}, {\mathbf{r}^l}'] 
    \coloneqq \Vert \mathbf{p}_{\mathbf{r}^l}^i - \mathbf{p}_{{\mathbf{r}^l}'}^i \Vert_{1}.
\end{equation}

\subsubsection{Step 3: Adjacency Matrix}
\noindent
Next, we construct the adjacency matrix, which captures the temporal and spatial relationships among patches.
We use the \ac{knn} \cite{zhang2016introduction} to cluster the patches based on their similarities and form the adjacency matrix, $\mathbf{M}$, as
\begin{equation}
\label{eq:l1distbasedAdjacency}
    \mathbf{M} \coloneqq {\tt{KNN}_\alpha} \left({\tt GeLU} 
    \left({\mathbf{D}}\right)\right),
\end{equation}
where $\alpha$ is {\tt KNN} sparsity factor, and {\tt GeLU} is the Gaussian error linear unit.

\subsection{Proposed Model} 
\noindent
Given the patches and adjacency matrix, the proposed hybrid \ac{tt} model first performs patch embedding and positional encoding, followed by a \ac{sda} operation to discover periods in the frequency domain that convert the embedded patches into a 2D time series.
Then, the model uses a parameter-inception block to learn extremely intricate patterns in the frequency domain, followed by period aggregation to convert the 2D series into a 1D time series. 
To learn relevant correlations among patches in the graph, we  divide the graph into multiple ego-graphs and then apply a \acp{moe} allocation \cite{hu2025timefilter}. 
Finally, we reconstruct a sparse graph to form a global graph, which is used for (\ac{gnn}-based) training.
These steps are summarized below.

\subsubsection{Patch Embedding}
\noindent
In order to construct a learnable graph, we first embed the patches in $\mathbf{p}_{\mathbf{r}^l}$ to a higher dimensional, $C$, feature space.
In the training graph, each embedded patch is essentially a node, and flattening the output after the embedding gives a shape of $N \coloneqq{C}\times {I}$,
where $I=7$ is the total number of patches at each timestep.
\begin{equation}
    \mathbf{X}_{\mathrm{embed}}= {\tt Embedding}
(\mathbf{p}_{\mathbf{r}^l}^i) \in\mathbb{R}^{ N\times \lfloor d\rfloor},
\end{equation}
where $d$ is the embedding size.

To this end, we use a multihead operation \cite{hu2025timefilter} to segment $\mathbf{X}_{\mathrm{embed}}$ into $H$ heads. 
Denote head $h$ by $\mathbf{X}_{{h}} \in\mathbb{R}^{H\times N\times \lfloor d/H \rfloor}$.
We then compute the corresponding distance matrix as
\begin{equation}
    \mathrm{distance}(\mathbf{X}_{h})\coloneqq
\mathrm{Linear}(\mathbf{X}_{{h}})
*
\mathrm{Linear}(\mathbf{X}_{{h}})^{T}
\in
\mathbb{R}^{H\times N\times N}.
\label{eq:projection_distance}
\end{equation}
To learn the most relevant correlations, we apply a mask, $m_\alpha$, to $\mathrm{distance}(\mathbf{X}_{h})$ and construct a learnable adjacency matrix as
\begin{equation}
    \tilde{\mathbf{M}} \coloneqq {\tt{m}_\alpha} \left({\tt GeLU} 
    \left({\mathbf{\mathrm{distance}(\mathbf{X}_{h})}}\right)\right).
\end{equation}

\subsubsection{Combined Trainable Adjacency Matrix}
\noindent
The model then combines the trainable adjacency matrix $\tilde{\mathbf{M}}$ with \eqref{eq:l1distbasedAdjacency} to construct the final adjacency matrix as
\begin{equation}
\mathbf{M_f} \coloneqq {\tt softmax}(\mathbf{M}) + {\tt softmax}(\tilde{\mathbf{M}}).
\end{equation}

\subsubsection{Spectral Domain Analysis}
Since the combined adjacency $\mathbf{M_f}$ is in the time domain, intricate temporal patterns are difficult to learn.
To solve this, we use \ac{fft} to convert each embedded patch, analyzed in the time domain (1D), into the frequency domain (2D) to identify periods and learn intricate patterns \cite{wu2022timesnet}.
We then use a parameter-inception block to learn the intricate patterns using 2D kernels. 
After learning, we truncate the 2D time series into a 1D time series for residual summation to preserve its original form.
This is achieved through period aggregation, in which the frequency components from the FFT are aggregated into a 1D series.

\subsubsection{ Construction of ego-graphs and subgraphs}


Using the above steps, the proposed method now constructs a learnable graph $\mathcal{E}$ that may still contain irrelevant and noisy information.
To find the relevant information, we followed the steps proposed in \cite{hu2025timefilter}.
More specifically, the graph is divided into $Q$ ego graphs, $\{\mathcal{E}_t\}_{t=1}^Q$.
Each ego graph is further divided into temporal, spatial, and spatio-temporal graphs to learn different types of dependencies since the embedded patches could exhibit different dependencies, i.e., spatial, temporal, and spatio-temporal.
Then we use \acp{moe} to find the best filters for each embedded patch, i.e., node of the graph.
Next, these filters are used to filter the embedded patches and reconstruct a sparse graph, which is then leveraged to train our proposed \ac{tt} model using the \ac{gnn} training strategy.

\subsubsection{Fusion}
\noindent
To learn the spatial and temporal relationships in the graph, our \ac{gnn}-based architecture leverages message passing to capture these dependencies. 
We have a graph-level prediction task, i.e., the new representation of each node is the weighted sum of its neighborhood edges, and the new representation of each edge is the weighted sum of its neighborhood nodes. 
All these representations are combined to form the graph-level prediction, denoted by $\mathbf{X}_{p}$.
As described above, the parameter inception block is used to learn temporal patterns in 2D tensors and is then truncated back to 1D. 
The final output of this process is denoted as $\mathbf{H}_a$. To boost model performance, we combined the outputs as
\begin{equation}
\mathbf{Y}\coloneqq\mathbf{X}_\mathrm{embed}+\mathbf{H}_a+\mathbf{X}_{p}.
\end{equation}

\subsubsection{Layer Normalization}
The model then uses layer normalization as $\mathbf{Y} = {\tt LayerNorm} \left(\mathbf{Y}\right)$ to stabilize the training.

\subsubsection{Prediction head} 
The feedforward network generates the predictions as $\mathbf{Y} = \mathbf{Y} + {\tt GeLu}\left(\mathbf{W}_{1} \mathbf{Y} + \mathbf{b}_{1}\right)\mathrm{W}_2  + \mathbf{b}_2$.

\subsection{Statistics-aided Training Method} 
\label{sec:statAidedTraing}
\subsubsection{Statistics-aided Loss Function}
To facilitate the training process, we directly incorporate important propagation knowledge through the following customized loss function.
\begin{equation}
\label{eq:lossfun}
    \phi(s,\hat{s},g, \hat{g})\coloneqq\phi(s,\hat{s}) + \phi(g,\hat{g})  + \lambda_1 \mathcal{L}_{\text{dyn}} + \lambda_2 \mathcal{L}_{\text{imp}} , 
\end{equation}
where $\phi(s,\hat{s}) \coloneqq
\alpha_{\tau}\left(s_{\tau} - \hat{s}_{\tau}\right)^2
+ \alpha_{\Psi_{\mathrm{az}}}\left(s_{\Psi_{\mathrm{az}}} - \hat{s}_{\Psi_{\mathrm{az}}}\right)^2 + \alpha_{\Psi_{\mathrm{zn}}}\left(s_{\Psi_{\mathrm{zn}}} - \hat{s}_{\Psi_{\mathrm{zn}}}\right)^2 + \alpha_{\Omega_{\mathrm{az}}}(s_{\Omega_{\mathrm{az}}} - \hat{s}_{\Omega_{\mathrm{az}}})^2 + \alpha_{\Omega_{\mathrm{zn}}}(s_{\Omega_{\mathrm{zn}}} - \hat{s}_{\Omega_{\mathrm{zn}}})^2$ and $\phi(g,\hat{g})\coloneqq\alpha_g\!\sum_{n=1}^{n=M}\!\left(g_n\ - \hat{g}_n\!\right)^2$.
Besides, $\alpha_{*}$ are the weights corresponding to the statistics and gain.
Moreover, $\lambda_1 \mathcal{L}_{\text{dyn}}$  and  $\lambda_2 \mathcal{L}_{\text{imp}}$  are as defined in \cite{hu2025timefilter}.

\subsubsection{Back Propagation}
In order to train the model, we use mini-batch \ac{sgd}. 
More specifically, we get the loss using \eqref{eq:lossfun} during the forward propagation and then update the model using the backpropagation as 
\begin{equation}
    \pmb{\Theta} \gets \pmb{\Theta} - \eta \nabla_{\pmb{\Theta}} \big[\phi(s,\hat{s},g, \hat{g})\big],
\end{equation}
where $\pmb{\Theta}$ is the model parameters, $\eta$ is the learning rate, and $\nabla$ is the gradient operator.

\begin{table}[!t]
\centering
\caption{NMSE [in dB] between Statistics Calculated from top-$M$ Generated Channel Realizations $\&$ Ground Truth on \ac{scm} Dataset using the Proposed Solution}
\label{tab:impactofM}
\begin{tabular}{|c|c|c|c|c|}
\hline
\textbf{Name} & \textbf{M=5} &\textbf{M=10}  & \textbf{M=15}   \\
\hline
\textbf{DS} & -8.8030 &  -8.9474 & -8.9455   \\
\hline
\textbf{Az (AoA) AS}&    -8.4890 &     -8.5834 &   -8.5673 \\
\hline
\textbf{Az (AoD) AS} &        -8.4657 &        -8.5928 &     -8.5651 \\
\hline
\textbf{Zn (AoA) AS} &    -4.4881&  -4.9317  &   -4.9651 \\
\hline
\textbf{Zn (AoD) AS} &  -4.4667 &   -4.9444   &   -4.9808 \\
\hline
\textbf{Gain} &    -24.8145  &    -26.2784 &  -26.9663\\
\hline
\end{tabular}
\end{table}

\section{Simulation and Parameter configuration}

\subsection{Datasets Generation}
\noindent
For performance evaluation, we first generate (a) synthetic datasets based on \ac{scm} and (b) deterministic ray tracing datasets.
For the \ac{scm} dataset, we first generate receiver trajectory points using
$q_{\mathbf{r}^l,x} = \left(R + Z \sin\left(6\pi l/V\right)\right) \cos\left(2\pi l/V\right)$, 
$q_{\mathbf{r}^l,y} = \left(R + Z \sin\left(6\pi l/V\right)\right)\sin\left(2\pi l/V\right)$, $q_{\mathbf{r}^l,z} = 1.5$,
where $f = 0,1,\dots,V$, $V$ is the number of samples, $R=30$ is the base radius, and $Z=40$ is the radial  amplitude. 
The \ac{mpc} are randomly distributed using a {\em Poisson} point process with $0.0004$ intensity. 
We used the approach of  \cite{pervej2025double} to calculate the \ac{mpc} features.

\begin{figure*}
\begin{minipage}{\textwidth}
\centering
\includegraphics[width=\textwidth, height=0.25\textheight]{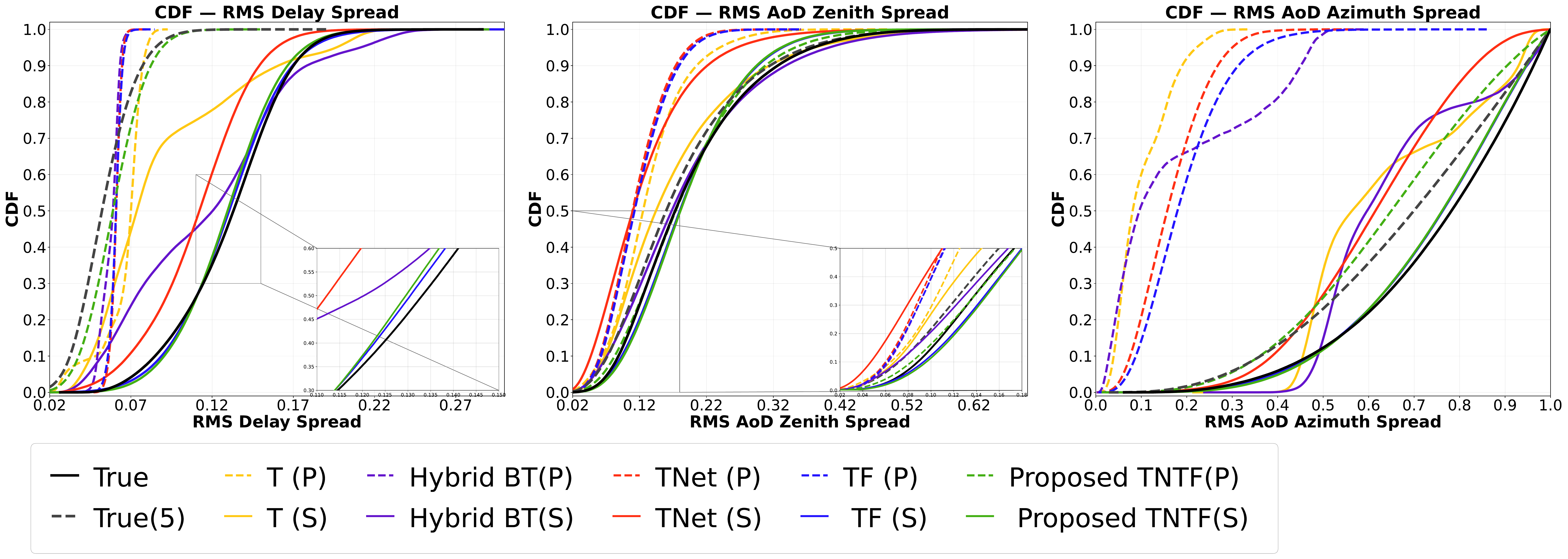}
\caption{CDF of the Statistics on SCM Datasets: $L=100$, $P=300$, and $M=5$}
\label{fig:2}
\end{minipage}
\begin{minipage}{\textwidth}
\centering
\includegraphics[width=\textwidth, height=0.2\textheight]{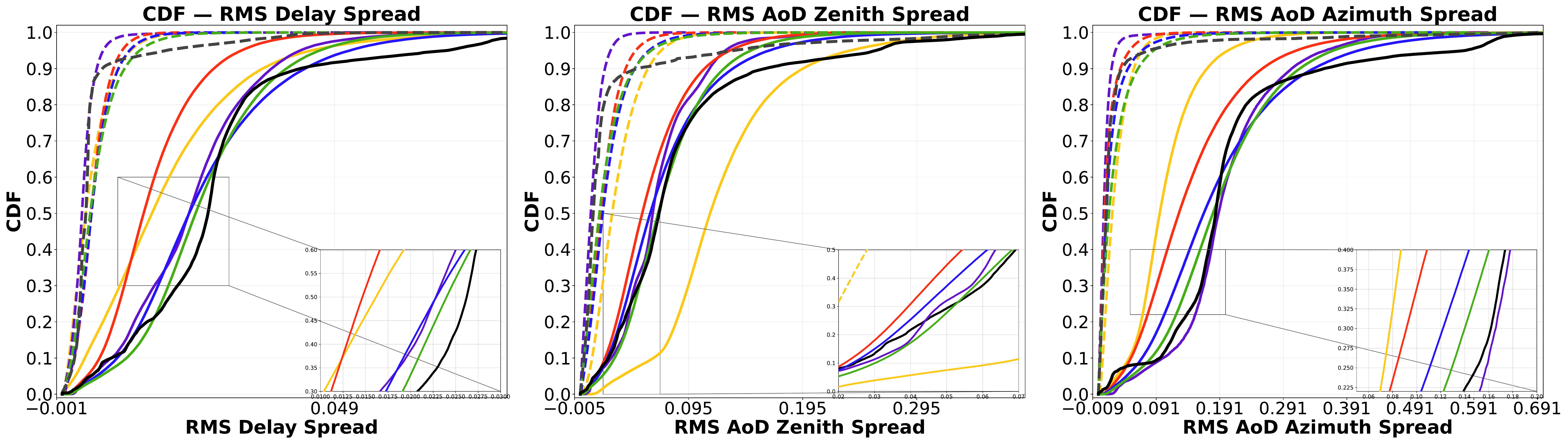}
\caption{CDF of the Statistics on Ray Tracing Dataset: $L=100$, $P=300$, and $M=2$}
\label{fig:3}
\end{minipage}
\end{figure*}

For the ray tracing dataset, we used OpenStreetMap, Blender, and Sionna\footnote{\url{https://nvlabs.github.io/sionna/rt/index.html}} to construct the \ac{dd} channels near Utah State University, Logan, UT, USA, based on predefined trajectory points.
We consider two types of trajectory generation functions.
For the first type, we have
$h(l) \coloneqq U + \alpha_1 l + n_h(l)$,
$q_{x,\mathbf{r}^l} = w + \rho h(l)\cos(\theta(l))$, 
$q_{y,\mathbf{r}^l} = l + \lambda h(l)\sin(\theta(l))$, 
$q_{z,\mathbf{r}^l} = m + n_z(l)$, 
where $h(l)$ is the spiral radi. 
$n_h(l), n_\theta(l)$, and $n_z(l)$ are Gaussian noises.
Besides, $U=\{5,8,6\}$, $w=\{45,48,42\}$, $l=\{70,68,72\}$, $m=\{1.5,1.8,1.2\}$, $\rho=\{1,1,1.3\}$, and $\lambda=\{1,1,0.8\}$ for the $3$ trajectories, respectively.
Furthermore, $\theta(l)$ is the angular rotation, which is defined as $\theta(l) \coloneqq l + n_\theta(l)$ for the first and third trajectories, and $\theta(l) \coloneqq -l + n_\theta(l)$ for the second trajectory. 
Moreover, for the second type trajectory, we use $\theta(l) \coloneqq z + n_\theta(l)$, 
$q_{x,\mathbf{r}^l} = 45 + 12\sin(2\theta(l))$, 
$q_{y,\mathbf{r}^l} = 70 + 10\sin(\theta(l))$, and
$q_{z,\mathbf{r}^l} = 1.5 + 0.8\sin(l/2) + n_z(l)$.
Finally, we use $0~\mathrm{dBm}$ transmission power, $2.4~\mathrm{GHz}$ carrier frequency, and dipole antennas.

We generated $1$ Million samples, out of which $75 \%$, $5 \%$, and $25\%$ are used for training, validation, and test purposes, respectively.
Besides, we used the custom data normalization process from \cite{pervej2025double}.

\begin{figure*}
\centering
\begin{minipage}{\textwidth}
\includegraphics[width=\textwidth,height=0.2\textheight]{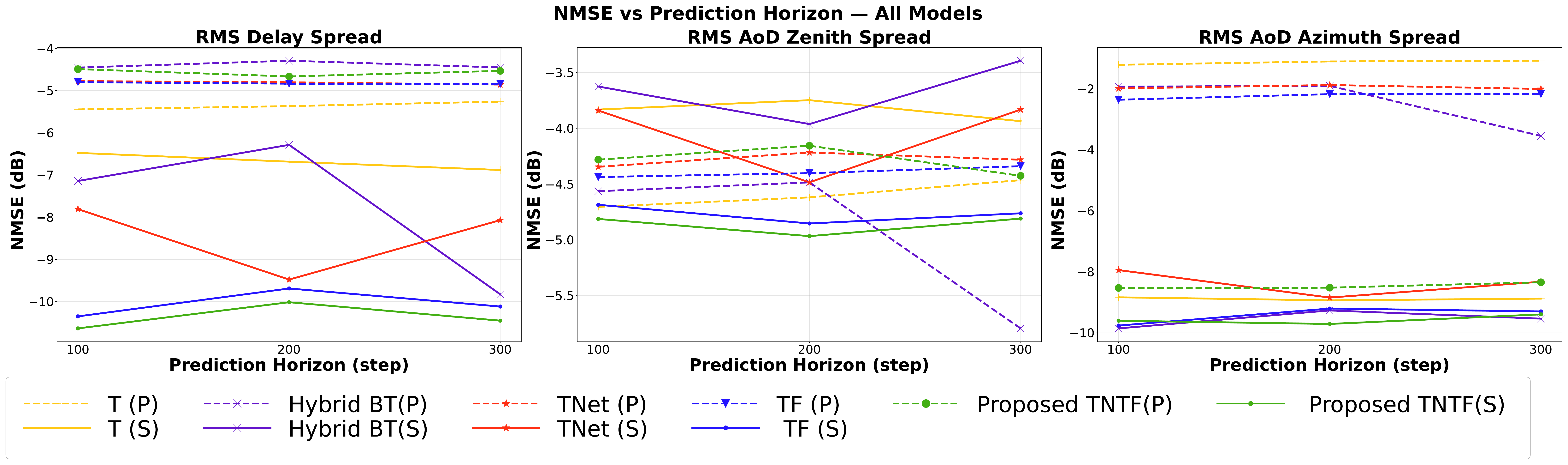}
\caption{NMSE [in dB] for different $P$ on SCM Dataset: $L=100$ and $M=5$}
\label{fig:4}
\end{minipage}
\begin{minipage}{\textwidth}
\centering
\includegraphics[width=\textwidth,height=0.2\textheight]{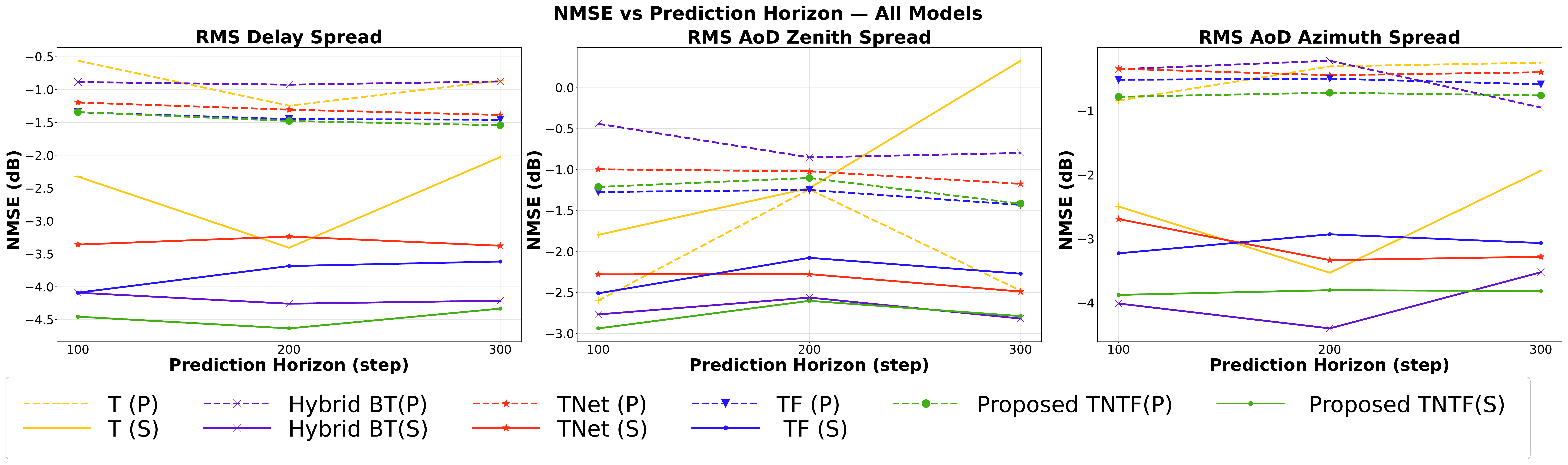}
\caption{NMSE [in dB] for different $P$ on Ray Tracing Dataset: $L=100$ and $M=2$}
\label{fig:5}
\end{minipage}
\end{figure*}

\subsection{Baselines and Hyperparameters}

\subsubsection{Baselines}
To evaluate the performance of our proposed solution, we consider several Transformer- and hybrid-architecture-based baselines.
Besides, for apples-to-apples comparisons, we used both regular loss functions, termed predictive (P), and our statistics-aided loss functions, termed statistics (S), to train and evaluate these models.
The considered baselines are: Transformer (T)-based \cite{vaswani2017attention} T(P) and T(S); hybrid BiLSTM-Transformer (BT)-based \cite{pervej2025double,pervej2026dd} BT(P) and BT(S); TimesNet (TNet)-based \cite{wu2022timesnet} TNet(P) and TNet(S); and TimeFilter (TF)-based \cite{hu2025timefilter} TF(P) and TF(S).
Moreover, to show the effectiveness of our proposed \ac{tt} model, we consider both the regular loss (\ac{tt}(P)) and statistics-aided loss (\ac{tt}(S)). 

\subsubsection{Hyperparameters}

We use ablation study to find the best hyperparameters.
For the TNTF(P), TNTF(S), TF(S), TF(P), TNET(P), and TNET(S) models, we use the following hyperparameters on \ac{scm} dataset: $H=2$, $\mathrm{d}_{\mathrm{model}} =8$, $2$ layers (of timeblocks for TNET and graph blocks for TF-based models), $16$ $FC$ layers,  $\alpha=0.1$, dynamic expert factor of $0.5$, and batch size of $32$. 
For the \ac{tt} model, \ac{fft} period, kernel, and $\eta$ are $2,2$ and $0.0003$, respectively.
For the TNET(P) and TNET(S), we use $\mathrm{d}_{\mathrm{model}}$, \ac{fft} period, and kernel of $16$, $3$, and $3$, respectively.
For the BT(P), BT(S), T(S), and T(P) models, we use $H=2$, $\mathrm{d}_{\mathrm{model}} =32$, $2$ encoder/decoder layers, $32$ FC layers. 
Besides, we use $1$ LSTM layer with $16$ hidden dimension for the BT(P) and BT(S) model.
Furthermore, we used batch size of $256$ and $128$ for the BT(S)/BT(P) and T(P)/T(S), baselines.

In the ray tracing dataset, we used $\mathrm{d}_{\mathrm{model}} =8$, $2$ layers, FC layer with $8$ neurons, and batch size of $32$ for the TNTF(S), TF(S), TNET(S), TNET(P), and TF(P) models.
Besides, the TNTF(P), TNTF(S), TF(S), and TNET(S) models use $H=4$, , $\alpha=0.4$, dynamic expert factor $=0.5$, \ac{fft} period of $9$,  and kernel of $9$. 
The TNET(P) and TNET(S) models used  $\eta=0.0001$, $H=2$, $\mathrm{d}_{\mathrm{model}} =8$, periods of $9$ and kernel of $9$. 
The BT(P), BT(S) used batch size of $128$, $\eta=0.0001$, $H=8$ , $\mathrm{d}_{\mathrm{model}} =8$,  2 layers,  $16$ $FC$ layers, $1$ LSTM layer with hidden dimension of $8$.
The T(P) and T(S) used batch size of $256$, $\eta=0.0001$, $H=8$, $\mathrm{d}_{\mathrm{model}}=8$,  2 encoder/decoder layers, and $16$ $FC$ layers.
In all cases, we use the Adam weight (AdamW) optimizer and train the models for $250$ epochs.

Unless otherwise mentioned, the results reported below for all baselines are the {\em average} from {\em three independent trials} for the \ac{scm}-based synthetic dataset, which accounts for stochasticity. 
Besides, a single run is used to obtain results for the deterministic ray-tracing-based dataset. 

\subsection{Results and Discussions}

\subsubsection{Impact of $M$}
Intuitively, increasing the value of $M$ (for top-$M$ \acp{mpc}) should improve the model performance, as it provides more complete channel information.
However, in many cases, the \ac{dd} channel may have only a few representative \acp{mpc} that dominate the overall channel. 
We observe this effect in Table \ref{tab:impactofM}, which shows that as $M$ increases, the \ac{nmse} decreases, though the decrease may not be substantial. 

\subsubsection{Predictive vs. Proposed Statistics-Aided Solution}
Since we are only using partial channel information, typical predictive solutions are likely to fail when the top-M MPCs do not dominate.
This is because predictive solutions will only try to match the top-M MPCs, thereby largely deviating from the actual statistics.
In contrast, the statistics-aided solution shall still work well, since it learns to directly match the statistics using the training method summarized in Section \ref{sec:statAidedTraing}.


These trends are also observed in our simulation results on both datasets. Fig. \ref{fig:2} shows the trend for the SCM dataset with $M=5$: the statistics-aided solutions work better than the predictive solutions. 
Among the statistics-aided baselines, the proposed TNTF(S) has the most closely aligned \ac{cdf} with the ground-truth statistics. 
Fig. \ref{fig:3} shows the trend for the ray tracing dataset with $M=2$.
Even in this extreme case, our proposed TNTF(S) demonstrates a closely aligned \ac{cdf} for the considered channel statistics.

\subsubsection{Impact of Future Sequence Length $P$}
When the prediction length $P$ increases with a fixed $M$, the predictive solutions may be unreliable if channel realizations change frequently and there are clearly more dominant MPCs than $M$.
This may also create problems for the statistics-aided solution. 
However, since channel statistics are less likely to change rapidly, a statistics-aided solution shall still be a clearly better choice. 
Our simulation results for various $P$ are shown in Fig.~\ref{fig:4} and Fig.~\ref{fig:5}  for the SCM and ray tracing datasets, respectively. 
Clearly, we observe that the statistics-aided training method is more stable than most statistics-aided baselines across both datasets.

\section{Conclusions}
\noindent
This work proposed a novel way for DD propagation channel modeling using a statistics-aided GNN-based solution. 
We used a subset of MPCs to mitigate the problem of varying numbers of MPCs across locations, then constructed a learnable graph using propagation-knowledge-informed patching, followed by modeling temporal, spatial, and tempo-spatial correlations among the embedded patches with our proposed TNTF model. 
To this end, we train the model using a statistics-aided training method to retain domain knowledge and mitigate incomplete channel information. 
Our extensive simulations on SCM and ray-tracing datasets demonstrated the effectiveness of the proposed solution.

\vspace{-0.05in}
\bibliographystyle{IEEEtran}
\bibliography{reference}

\end{document}